\documentclass[twocolumn,printnumbers,amsmath,amssymb,showpacs]{revtex4}

\newcommand{\be}{\begin{equation}}
\newcommand{\ee}{\end{equation}}
\newcommand{\ba}{\begin{eqnarray}}
\newcommand{\ea}{\end{eqnarray}}
\usepackage{graphicx}
\usepackage{dcolumn}
\usepackage{bm}

\begin{document}

\title{Heat transport in model jammed solids}


\author{Vincenzo Vitelli$^1$} 
\author{Ning Xu$^{1,2}$} 
\author{Matthieu Wyart$^{3}$} 
\author{Andrea J. Liu$^1$}
\author{Sidney R. Nagel$^2$}

\affiliation{$^1$Department of Physics and Astronomy, University of
  Pennsylvania, Philadelphia PA, 19104; $^2$The James Frank Institute, The
  University of Chicago, Chicago IL, 60637; $^3$HSEAS, Harvard University,
  Cambridge MA, 02138}

\begin{abstract}
We calculate numerically the normal modes of vibrations in 3D jammed packings of soft spheres as a function of the packing fraction and obtain the energy diffusivity, a spectral measure of transport that controls sound propagation and thermal conductivity.  
The crossover frequency between weak and strong phonon scattering is controlled by the coordination and shifts to zero as the system is decompressed towards the critical packing fraction at which rigidity is lost. Below the crossover, the diffusivity displays a power-law divergence with inverse frequency, which suggests that the vibrational modes are primarily transverse waves, weakly scattered by disorder. Above it, a large number of modes appear whose diffusivity plateaus at a nearly constant value independent of the inter-particle potential, before dropping to zero above the Anderson localization frequency. The thermal conductivity of a marginally jammed solid just above the rigidity threshold is calculated and related to the one measured experimentally at room temperature for most glasses.  
\end{abstract}

\maketitle

\section{\label{sec:intro}Introduction}

The thermal and mechanical properties of disordered solids can differ dramatically from those of
crystalline materials \cite{Elliott_book}.  Prominent among the anomalous properties are the specific heat and thermal conductivity, which display common features at sub-helium to room temperatures, in amorphous materials ranging from 
glasses to plastics and even frozen grease \cite{Phil81}.  This commonality suggests that the explanation of the unusual features
of disordered solids may involve general physical principles that transcend detailed information about the 
chemical structure of specific compounds \cite{Lub03,Sheng91,Allen93}.  

This paper focuses on the {\it intermediate} temperature regime, $1K<T<T_{\rm room}$.    In this regime, the thermal conductivity has a plateau followed by a nearly linear rise at higher $T$, in contrast to the sharp drop seen in crystalline materials in the same range~\cite{Pohl87}.  In the temperature range of this thermal-conductivity plateau, the ratio of the heat capacity $C(T)$ to the expected $T^3$-dependence predicted by the Debye model for crystalline solids, $C(T)/T^3$, exhibits a prominent peak, termed the boson peak, which is a hallmark of amorphous solids \cite{Phil81, XLiu}.  

At lower temperatures, $T<1K$, the thermal conductivity exhibits a $T^2$-rise, in contrast to the $T^3$ dependence observed in crystals \cite{zp71}.  Meanwhile the heat capacity increases linearly in $T$ \cite{Phil81}, in contrast to the $T^3$ rise predicted by Debye for long-wavelength sound modes.  These low-temperature features have been rationalized by invoking the scattering of long-wavelength phonons off two-level systems, 
posited to arise from groups of atoms tunneling between
two minima \cite{Pohl02,2lev}.  While this view of the low-temperature regime is widely accepted, little consensus exists regarding the intermediate temperature regime.  In particular, the origin of the plateau in the thermal conductivity is controversial, and the question of whether the plateau in the thermal conductivity is linked to the boson peak remains unresolved.

In this article we present a systematic study of the pressure (or packing fraction) dependence of the energy diffusivity of random packings of soft frictionless spheres. The energy diffusivity, $d(\omega)$, quantifies how far a wavepacket narrowly peaked at a frequency $\omega$ can propagate.  We study a class of models in which spheres interact via a repulsive potential that vanishes at a well-defined distance that defines their diameter.  These models possess a jamming/unjamming transition as a function of packing density~\cite{O'Hern03}.  Thus, the strength of the elastic moduli can be tuned continuously downwards by decreasing the density of the packing until the particles no longer interact.  At that critical density, the disordered solid unjams to form a liquid.  As a result, the onset of the excess vibrational modes $\omega^{*}$ can be pushed  to zero by decreasing the packing fraction \cite{Leo1}. Numerical calculation of the density of states and the diffusivity enables us to track the pressure dependences of the boson peak frequency $\omega^{*}$ and the transport crossover frequency $\omega_{d}$ individually and to compare them.

We find that the diffusivity \cite{Xu09} displays a well-defined kink at a frequency $\omega_{d}$ that separates the low $\omega$ regime of divergent diffusivity from
a characteristic plateau that persists up to high $\omega$ where Anderson localization sets in.  Moreover,  $\omega_{d}$ and the boson peak frequency, $\omega^*$, are not only comparable in magnitude, but decrease in tandem as we decrease packing fraction towards the unjamming transition.  This shows that the transport crossover is linked to the excess modes.  

A unique feature of disordered sphere packings is that the transport crossover can be understood perturbatively by means of simple scaling laws parameterized in terms of the distance from the critical unjamming transition, denoted as Point J \cite{O'Hern03}.  Furthermore, some of the distinctive properties of amorphous solids are manifested in their most extreme form as the unjamming transition is approached.  For example, in that limit the diffusivity plateau extends all the way to the lowest frequencies studied in the simulations.  This suggests that the origin of the diffusivity plateau can be traced to properties of the unjamming transition \cite{Xu09}. 

The outline of this paper is as follows. In Section \ref{sec:back}, we provide the necessary background on the link between vibrational dynamics and heat transport on which our work is built and review the scaling properties of jammed solids. In Section \ref{meth}, we review the methodology adopted to calculate the energy diffusivity using the Kubo formula, which enables a first-principles calculation of the diffusivity for computer-generated packings. In section \ref{results}, we present our results for the transport crossover and show that this occurs at the boson peak frequency.  We also present a scaling analysis that rationalizes how the main features of the diffusivity depend on the distance from the unjamming transition for both Hertzian and harmonic interactions. Section \ref{sec:plateau} focuses on the most striking transport signature of jammed packings: a plateau in the diffusivity as a function of frequency.  Such a plateau was postulated by Kittel~\cite{Kittel} in order to explain the temperature dependence of the thermal conductivity of amorphous solids at intermediate temperatures; our results provide evidence that this scenario is correct.  We explain the origin of the diffusivity plateau, 
starting from a set of assumptions concerning the nature of the vibrational modes above the transport crossover. The AC thermal conductivity at point J is then obtained in section \ref{ac}. In Section \ref{conclusion} we conclude by summarizing the broader message of this article: the thermal conductivity of various amorphous materials under pressure can be explained from the vibrational modes at Point J, which controls energy transport at higher densities in the manner expected for a critical point.

\section{\label{sec:back}Background}  

It was originally suggested by Kittel \cite{Kittel} that the intermediate temperature properties might be understood from the microscopic vibrational dynamics of amorphous materials. 
For instance, the boson peak at $T^{*}$ implies the presence of vibrational modes in the density of states at $\omega^*=k_B T^*/\hbar$ in excess of the usual Debye counting.  Similarly, the plateau in the thermal conductivity could result from a crossover in the behavior of the mean free path of phonons \cite{anderson}.  

Several theoretical models have been advanced to explain the boson peak, starting from distinct physical mechanisms such as the existence of resonant (quasi)-localized modes \cite{ls91,so96,cjj96,Biswas88}, anharmonic interactions induced by the presence of defects \cite{bgg92,kki83}, the breakdown of continuum elasticity below a characteristic length scale \cite{matthieu,Wit02} or quenched disorder in the elastic constants \cite{sch06,sch07,ell92}.  Some of these models find that the onset of the excess vibrational modes coincides with a crossover from weakly-scattered
plane waves to strongly-scattered vibrational modes that are delocalized and poorly conducting \cite{sch98,Parshin}, suggesting a connection between the boson peak and the plateau in the thermal conductivity.  

The connection between the boson peak and the transport crossover has been probed using Brillouin scattering measurements but no firm conclusion has been reached to date \cite{fcv96,mrs96}.   A recent experimental study suggested that the excess modes in the density of states appear around the Ioffe-Regel frequency \cite{monaco}, at which the mean free path of longitudinal phonons approaches their wavelength \cite{Ioffe}. This heuristic criterion to estimate the crossover between weak and strong scattering was also used in an independent study that challenges the previous claim by concluding instead that the Ioffe-Regel limit takes place at a frequency significantly higher than the Boson peak frequency \cite{scopigno}. Classic experimental studies of Raman scattering point to another important piece of experimental evidence, namely that the vibrational modes at the Boson peak are transverse in character \cite{winterling,nemanich}. This observation is supported by simulations of silica \cite{pilla,horbach} and soft sphere glasses \cite{schober}.  Recent numerical simulations provided additional evidence that suggests the equality of the Boson peak frequency with the Ioffe-Regel limit for transverse phonons \cite{shintani}. 

\subsection{\label{sec:vibrations}Vibrational dynamics and heat transport}   

Instead of focusing directly on the heat capacity, $C(T)$, and thermal conductivity, $\kappa(T)$, we consider the density of states, $D(\omega)$, and diffusivity, $d(\omega)$.  
A heuristic argument for the relation between $d(\omega)$ and $\kappa(T)$ is as follows.  For a system in a temperature gradient, it is well known that the heat diffusivity obeys the relation $d=\kappa V/C$.  Thus, $\kappa=d C/V$.  This relation can be generalized mode by mode.  Thus, the heat capacity is \cite{John,Sheng91,Allen93}
\begin{equation}
C(T)= \sum_i C(\omega_i,T)  \label{eq:discretecapacity}
\end{equation}
where the sum runs over all vibrational modes $i$ and $C(\omega_i,T)$ is the heat capacity per mode, which is obtained from the Bose-Einstein distribution and is a universal function that characterizes the heat carried by a mode of frequency $\omega_i$ at temperature $T$.  Similarly, the thermal conductivity is
\begin{equation}
\kappa(T)=\frac{1}{V} \sum_i d(\omega_i) \!\!\!\!\! \quad C(\omega_i),  \label{eq:diffusion2}
\end{equation}
where $d(\omega_i)$ is the energy diffusivity of mode $i$.

We may recast Eqs.~\ref{eq:discretecapacity}-\ref{eq:diffusion2} in continuum form using the density of vibrational states:  
\begin{eqnarray}
C(T)&=& \int_{0}^{\infty} {\rm d}\omega \!\!\!\!\! \quad  D(\omega) \!\!\!\!\!  \quad C(\omega,T)   \!\!\!\!\! \quad , \label{eq:capacity} \\
\kappa(T)&=&\frac{1}{V} \int_{0}^{\infty} {\rm d}\omega \!\!\!\!\! \quad  D(\omega) \!\!\!\!\!  \quad d(\omega) \!\!\!\!\!  \quad C(\omega,T)  \!\!\!\!\! \quad . 
\label{eq:conductivity}
\end{eqnarray}
Both $D(\omega)$ and $d(\omega)$ are strongly {\it structure-dependent}: the density of states and diffusivity are the fingerprints of the vibrational modes in the solid and control its heat capacity and thermal conductivity.  

Inspection of Eq. (\ref{eq:capacity}) reveals that the prominent boson peak observed at the temperature $T^*$ in most amorphous solids is triggered by a large number of excess vibrational modes that show up in the density of states at a characteristic frequency $\omega^{*} \sim \frac {K_B T^*}{\hbar}$. It is known empirically that $\omega^*$ increases as the sample is compressed \cite{Gurevich,Hemley},  a property shared by the soft sphere packings investigated in this study.  By analogy, Eq. (\ref{eq:conductivity}) suggests that the origin of the thermal conductivity plateau around $T^{*}$ can be similarly traced to the existence of a transport crossover in $d(\omega)$ at $\omega^{*}$. 

At very low frequency, the diffusivity can be factored out as the product of the speed of sound $v$ and $\ell(\omega)$ \cite{Phil81}:
\begin{equation}
d(\omega)=\frac{v}{3} \!\!\!\!\! \quad  \ell(\omega)  \!\!\!\!\! \quad .
\label{eq:Cc}
\end{equation}
The mean free path $\ell(\omega)$ typically diverges as $\omega \rightarrow 0$ because the corresponding vibrational modes  can be regarded as long wavelength plane waves weakly scattered by disorder. However, once the mean free path shrinks to the order of the wavelength, Eq.~(\ref{eq:Cc}) breaks down \cite{Ioffe,Sheng-book}.
\subsection{\label{sec:packings} Jammed sphere packings}  
  
We study a model of amorphous packings of frictionless spheres
interacting via the repulsive pair potential $V(r_{ij})$
\begin{eqnarray}
V(r_{ij})&=&\frac{\epsilon}{\alpha}(1-r_{ij}/\sigma_{ij})^{\alpha}  \quad \quad \!\!  \text{if} \quad r_{ij}<\sigma_{ij} \nonumber \\
V(r_{ij})&=& 0  \quad \quad \quad \quad \quad \quad \quad \quad \text{if}  \quad r_{ij}>\sigma_{ij} \quad \!\!\!  ,
\label{eq:C}
\end{eqnarray}
where the distance
between the centers of particles $i$ and $j$ is denoted by $r_{ij}$ and the sum of their radii by
$\sigma_{ij}$. We generate $T=0$ packings by conjugate-gradient energy minimization according to the procedure described in Ref. \cite{O'Hern03}. 
Irrespective of the value of $\alpha$,
this model system exhibits a jamming/unjamming transition at $T=0$ at a packing fraction $\phi=\phi_c$ (Point J) at which the particles are just touching each other and there is no overlap~\cite{O'Hern03}. 

The zero-temperature jamming/unjamming transition has mixed character.  At this transition, the average coordination number, $z$, jumps~\cite{Durian95,O'Hern03} from zero to the minimum value required for mechanical stability, the ``isostatic" value $z_c=2D$~\cite{alexander}, where $D$ is the dimensionality of the sample. At densities lower than $\phi_c$, particles are free to rearrange while above $\phi_c$ at $\Delta \phi \equiv \phi-\phi_c$ , the system behaves as a weakly-connected amorphous solid with an average coordination number that scales as a power law with an exponent consistent with $1/2$~\cite{Durian95,O'Hern03}:
\begin{eqnarray}
\Delta z \equiv z-z_c \sim \Delta \phi^{1/2} \!\!\!\!\! \quad . \label{eq:scaling00} 
\end{eqnarray}   
In addition, both elastic moduli exhibit scaling behavior near the jamming point consistent with~\cite{Durian95,O'Hern03}:
\begin{eqnarray}
&G& \sim \Delta \phi^{\alpha-3/2}  \!\!\!\!\! \quad , \label{eq:scaling0} \\
&B& \sim \Delta \phi^{ \alpha -2}  \!\!\! \quad .
\label{eq:scaling1}
\end{eqnarray}  
For harmonic repulsions ($\alpha=2$), the bulk modulus is independent of compression while the shear modulus vanishes as point J is approached.  The bulk modulus scales as the second derivative of the potential with respect to compression, while the scaling of the shear modulus does not follow this naive scaling.

The ratio $G/B \sim \Delta \phi^{0.5}$ of the two elastic moduli is independent of $\alpha$ and controls the relative contribution of transverse to longitudinal waves at low frequency. This can be checked explicitly by considering that the phonon density of states at very low frequency satisfies the ubiquitous Debye law $D(\omega)\sim \frac{\omega^2}{v^3}$ except at Point J where, as we shall see,
the Deybe regime is completely swamped by a new class of vibrational modes. The transverse and longitudinal speeds of sound $v_{t}$ and $v_{l}$ are proportional to the square root of the shear and bulk moduli, respectively  
\begin{eqnarray}
&v_t& \sim \Delta \phi^{\frac{2 \alpha -3}{4}}  \!\!\!\!\! \quad ,  \label{eq:scaling2bis}  \\
&v_l& \sim \Delta \phi^{\frac{ \alpha -2}{2}}  \!\!\! \quad .
\label{eq:scaling2}
\end{eqnarray}  
Upon substituting into the Deybe formula, $D(\omega) \sim \omega^2/v^3$, Equations (\ref{eq:scaling2bis}) and (\ref{eq:scaling2}) imply that the ratio of the transverse to the longitudinal density of states at low frequency, $D_t/D_l \sim \Delta \phi ^{-3/4}$, becomes arbitrarily large as $\Delta \phi  \rightarrow 0$.  Thus, the density of states is dominated by transverse modes at low frequencies where wave-like behavior is expected, irrespective of the potential. 

Numerical studies~\cite{O'Hern03,Leo1} have revealed the presence of excess vibrational modes that contribute to a plateau in the density of vibrational states above a characteristic frequency, $\omega^*$. Close to the jamming point $\omega^*$ increases with density consistent with the power law  ~\cite{Leo1}
\begin{eqnarray}
\omega^{*} \sim \Delta \phi^{\frac{ \alpha -1}{2}}  \!\!\! \quad .
\label{eq:scaling3}
\end{eqnarray}
Thus, the plateau extends to zero frequency in a marginally-jammed solid ({\it i.e.} a packing of particles just above the onset of mechanical rigidity).   

\section{\label{meth} Methods and Model}

The energy diffusivity, $d(\omega)$, introduced in Sec.~\ref{sec:vibrations}, can be viewed physically in terms of the behavior of a spherically-symmetric wavepacket narrowly peaked at a frequency $\omega$ and localized
at position $\vec r$ at time $t=0$.  The spatial width of the wavepacket
spreads out because the normal mode components from which the wavepacket is constructed propagate in all directions.
The time-independent diffusivity, $d(\omega)$, is given by the square of
the width of the wavepacket at time $t$, divided by $t$ at long times $t$ \cite{Sheng91}. 

If the width grows linearly in time, the diffusivity is infinite; this corresponds to ballistic propagation.  If width grows with the square-root of time, a finite diffusivity is obtained; this corresponds to diffusive propagation.   As first noted by Anderson, a third possibility exists, namely that the width of the wave-packet saturates to a constant value over which the vibration is localized \cite{Anderson}.  Such localized modes, typically occur at high $\omega$ \cite{John}.  The diffusivity is vanishingly small and $d(\omega)$ cannot be factorized into the product of $\ell(\omega)$ and $v$, as in Eq. \ref{eq:C}, because no speed of sound can be associated with such vibrational modes. 

In this study, we calculate the diffusivity by evaluating the Kubo formula for $d(\omega)$ directly \cite{Allen93} for computer-generated packings in terms of the normal modes over the entire frequency range available. The rationale behind this choice is two-fold.  On one hand, we use the energy diffusivity as a spectral measure of transport to probe the character of the vibrational normal modes of a jammed solid. On the other hand, we use the jammed solid as a model amorphous structure whose transport properties can be studied as a function of pressure simply by varying the density relative to that of the unjamming transition.  This allows variation over orders of magnitude of pressure, which cannot be realized in more realistic models of molecular or network glasses. 

\subsection{\label{energy} Review of the Kubo formula for the energy diffusivity}

The analogy between the thermal conductivity in the phononic problem and its better-studied electrical counterpart underlies many of the mathematical techniques and physical concepts employed in our investigation. 

The Kubo formula for the energy diffusivity was derived in a convenient form by Allen and Feldman in Ref. \cite{Allen93}. Consider the $volume-averaged$ energy current $S$ that arises in response to an applied thermal gradient $\nabla T$. In linear response, $S$ is given by
\begin{equation}
S= -\kappa \!\!\!\!\! \quad \nabla T \!\!\!\!\! \quad ,
\label{eq:fourier}
\end{equation}     
where $\kappa$ is the thermal conductivity. More generally, the AC thermal conductivity $\kappa_{\mu \nu}(T,\Omega)$, which relates the energy flux $S_{\mu}$ in the $\mu$ direction to the time-varying temperature gradient in the $\nu$ direction, $\partial_{\nu} T \!\!\!\!\! \quad e^{i \Omega t}$, is
\begin{equation}
\kappa_{\mu \nu}(T,\Omega)= \frac{1}{VT} \int_{0} ^{\beta} d \lambda \int_{0} ^{\infty} dt \!\!\!\! \quad e^{i (\Omega + i \eta)t}  \left<S_{\mu}(-i \hbar \lambda) S_{\nu}(t) \right> \!\!\!\!\! \quad ,
\label{eq:kubo1}
\end{equation}     
where $V$ is the volume of the system and the angular brackets denote an equilibrium average of the autocorrelation function of the energy current operator $\hat{S}$. 

(Note that Eq. (\ref{eq:kubo1}) is analogous to the Einstein relation for the diffusion coefficient of an ensemble of Brownian particles in terms of the velocity-autocorrelation function. The main difference lies in the fact that in the thermal problem the conserved quantity is the energy density $\hat{h}$ that obeys the continuity 
equation: 
\begin{equation}
\frac{\partial \hat h}{\partial t}= - \vec{\nabla} \cdot  \vec{s}(r) \!\!\!\!\! \quad ,
\label{eq:kubo2}
\end{equation}     
where $s(r)$ is the local energy-flux operator (assumed to be isotropic for simplicity) prior to the spatial average that leads to $\hat{S}$.  In the case of the Einstein relation, the conserved quantity is the particle number and the role of $s(r)$ is played by the particle current.)

Our aim is to extract an expression for the diffusivity $d(\omega)$ by comparing Eq. (\ref{eq:kubo1}) to Eq. (\ref{eq:conductivity}) when the DC limit of $\kappa(T,\Omega)$ is taken, that is, when $\Omega \rightarrow 0$.  We first recast Eq.~(\ref{eq:kubo1}) in terms of a discrete sum over modes $i$:
\begin{equation}
\kappa_{\mu \nu}(T,\Omega)= \frac{1}{VT} \sum_{ij} \frac{n_{j}-n_{i}}{\hbar(\omega_i -\omega_j)} (S_{\mu})_{ij} (S_{\nu})_{ji} \delta(\omega_i - \omega_j - \Omega)   \!\!\!\!\! \quad ,
\label{eq:kubo3}
\end{equation}     
where $n_{i}$ is the equilibrium occupation number for bosons $n_{i}=[\exp(\beta \hbar \omega_i)-1]^{-1}$ and $(S_{\mu})_{ij}$ is the matrix element of the energy flux operator 
in the $\mu$ direction.  This matrix element can be computed from the vibrational normal modes, which are obtained from the dynamical matrix , $H_{\alpha\beta}^{mn}$, whose $i^{th}$
normalized eigenvector is denoted by $e_{i}(m; \alpha)$, where $\{m,n\}$ and $\{\alpha,\beta\}$ label particles and their Cartesian coordinates respectively \cite{Ashcroft}.  For disordered solids, the modes must be determined by numerical diagonalization of the dynamic matrix.

In an isotropic system a scalar thermal conductivity $\kappa(T)$ can be meaningfully defined from the trace of the tensor $\kappa_{\mu \nu}(T)$     
\begin{equation}
\kappa(T)= \frac{1}{3} \left( \kappa_{xx} + \kappa_{yy} + \kappa_{zz} \right)   \!\!\!\!\! \quad ,
\label{eq:kubo4}
\end{equation}     

To simplify Eq. (\ref{eq:kubo3}) further, consider that in the limit $\Omega \rightarrow 0$ the delta function forces the factor $(n_{j}-n_{i})/(\omega_i -\omega_j)$ to become $-\partial n_{i}/ \partial \omega_i$. One can then use Eq. (\ref{eq:kubo4}) in conjunction with the identity  
\begin{equation}
C(\omega_{i},T)=-\left(\frac{\hbar^2 {\omega_{i}}^2}{VT}\right) \!\!\!\!\! \quad \left(\frac{\partial n_{i}}{\partial \omega_i}\right) =k_B {(\beta \hbar\omega_i)}^2 \frac{e^{\beta\hbar\omega_i}}{{(e^{\beta \hbar\omega_i}-1)^2}}  \!\!\!\!\! \quad
\label{eq:clong}
\end{equation}   
to show that the thermal conductivity of Eq. (\ref{eq:kubo3}) can indeed be factorized as indicated in Eq. (\ref{eq:diffusion2}) with $d(\omega_{i})$ given by \cite{Allen93} 
\begin{equation}
d(\omega_{i})\equiv \frac{\pi}{3} \sum_{j} (\hbar \omega_{i})^{-2} \!\!\!\!\! \quad |\vec{S}_{ij}|^2 \!\!\!\!\! \quad \delta(\omega_i - \omega_j) ,
\label{eq:kubo5}
\end{equation}   
where the matrix elements $\vec{S}_{ij}$ read \cite{Allen93}
\begin{equation}
\vec{S}_{ij}= \frac{(\omega_i + \omega_j)^2}{4 \!\!\!\!\! \quad \omega_i  \!\!\!\!\! \quad \omega_j} \sum_{m n,\alpha \beta} (\vec{r}_{m}-\vec{r}_{n}) e_{i}(m; \alpha) \!\!\!\! \quad H_{\alpha\beta}^{mn} \!\!\!\! \quad e_{j}(n; \beta) \!\!\!\! \quad .
\label{eq:Sij}
\end{equation}

In the limit $\omega_i \rightarrow \omega_j$ enforced by the delta function in Eq. (\ref{eq:kubo5}), the prefactor $\frac{(\omega_i + \omega_j)^2}{4 \!\!\!\!\!\! \quad \omega_i  \!\!\!\!\!\! \quad \omega_j} \rightarrow 1$. However, taking this limit requires special care when the Kubo formula (derived in the continuum limit ) is evaluated for a finite and isolated system with a discrete spectrum \cite{Daniel}. Inspection of Eq. (\ref{eq:Sij}) reveals that the diagonal matrix elements $S_{ii}$ vanish. On the other hand any contribution to $d(\omega_{i})$ coming from the non-diagonal matrix elements $S_{ij}$ with $i \ne j$
is given zero weight when the delta function $\delta(\omega_i - \omega_j)$ is strictly enforced. This difficulty can be circumvented by smoothing out the $\delta$-functions in Eq. (\ref{eq:kubo5}) with the small finite width $\eta$. 
\begin{equation}
\delta _{\eta} (\omega_i-\omega_j)=\frac{\eta}{\pi[(\omega_{i}-\omega_{j})^2 + \eta^2]}\!\!\!\! \quad .
\label{eq:delta}
\end{equation}   

This heuristic procedure is expected to give the correct "bulk" result as long as the broadening $\eta$ is (a) much larger than the average level spacing, $\Delta$, and (b) much smaller than any characteristic frequency scale relevant to the problem.  In this paper, the broadening of the 
delta function $\eta$ in Eq. (\ref{eq:delta}) is typically chosen to be approximately five times larger than 
the average level spacing, $\Delta$.   We have verified that our numerical results do not depend on this choice as long as conditions (a) and (b) are met.  In the Landauer formulation of transport, it is not necessary to introduce the level broadening $\eta$ by hand because the inherent coupling of the system to the reservoirs plays an analogous role \cite{Akkermans}.   

One advantage of studying the energy diffusivity instead of the thermal conductivity is that $d(\omega)$ is finite at nonzero frequency when evaluated at the harmonic level.  By contrast, $\kappa(T)$ is generally infinite if anharmonic corrections are ignored.  This is because $d(\omega)$ in the integrand of Eq.~\ref{eq:conductivity} diverges too strongly at low $\omega$ due to phonons that are progressively less scattered with increasing wavelength. 

One main result of this paper will be that a marginally jammed solid is an exception to the rule that the thermal conductivity should diverge within the harmonic approximation.   However, as the system is compressed above Point J, $\kappa(T)$ again diverges, as in the standard case.

In order to cure this divergence, additional scattering mechanisms, beyond harmonic theory, are typically invoked resulting in an additional contribution to the diffusivity, $d_{c}(\omega)$. Upon adding $d_{c}(\omega)$ to the harmonic contribution, $d(\omega)$, (for example, as if they were two conductors in series \cite{book-cond}), one obtains the total diffusivity $d_{T}(\omega)$:
\begin{equation}
d_{T}(\omega)^{-1}= d(\omega)^{-1}+d_{c}(\omega)^{-1} \!\!\!\!\! \quad ,
\label{eq:series}
\end{equation}     
The cut-off contribution $d_{c}(\omega)$ necessarily dominates at low temperature, but much progress can be made in understanding 
the plateau in the thermal conductivity and the subsequent rise by studying the harmonic contribution, which dominates as $T$ increases.  

\subsection{\label{unstressed} Model}
Our simulations are carried out on jammed sphere packings, as described in Sec.~\ref{sec:packings}.  Specifically, we study a 50/50 bidisperse mixture comprised of $250 \le N \le 10,000$ frictionless spheres with a diameter ratio of 1.4, interacting with potentials described in Eq.~\ref{eq:C} with $\alpha=2$ and $\alpha=5/2$.
The packing fraction at
the onset of jamming, $\phi_c$, is characterized by the onset of a nonzero
pressure.  We determine $\phi_{c}$ and obtain $T=0$ configurations at
controlled $\Delta \phi \equiv \phi-\phi_c$ as in Ref. \cite{Leo1}.
For each configuration, we diagonalize its dynamical matrix and find the eigenvectors and the corresponding eigenfrequencies, which are measured in
units of $\sqrt{\epsilon/M\sigma^2}$, where $M$ is the particle mass \cite{Leo1}.  

We primarily consider an "unstressed" version of this model \cite{matthieuleo}, which is particularly tractable.  We use energy-minimized configurations obtained from numerically generated jammed sphere packings as described above.  We then replace the interaction potential, $V(r_{ij})$, between each pair of overlapping particles with an unstretched spring with the same stiffness, $V^{\prime \prime}(r_{ij}^{eq})$, where $r_{ij}^{eq}$ is the equilibrium distance between particles $i$ and $j$.  Note that since $r_{ij}^{eq}$ takes a different value between distinct particle pairs there will be a distribution in the local values of the elastic constants. On the other hand, the fact that all springs are unstretched guarantees that there are no net forces between particles in their equilibrium positions so that stable configurations for the stressed system are also stable in the unstressed one.  

The unstressed packings correspond to dropping terms depending on the first spatial derivative of the potential, $V^\prime$, in the dynamical matrix obtained from expanding the energy around the equilibrium position of the particles
\begin{equation}
\delta E=\frac{1}{2} \sum_{n,m}\biggl [V^{\prime\prime}(r_{nm}) (\delta \vec 
R_{nm}\cdot \hat r_{nm})^2  +  \frac{V^{\prime}(r_{nm})}{r_{nm}} (\delta \vec 
R_{nm}^\perp)^2  \biggr ].  \label{eq:Eexpn}
\end{equation}
where $n,m$ index particles.
 
The approximation of dropping the $V^\prime$ stress term in Eq.~\ref{eq:Eexpn} generates an interesting disordered system in its own right. The resulting off-lattice model, comprised of point particles interacting with relaxed springs, exhibits both spatial fluctuations in the local elastic stiffness as well as topological disorder (e.g., fluctuations in the local coordination number). Moreover, its amorphous structure can be varied by changing the volume.

The effect of the stress term can be seen from Eq. (\ref{eq:Eexpn}).  Since $V^\prime$ is negative for repulsive interactions, the stress term lowers $\delta E$ and hence the mode frequency.  We will discuss the effect of the stress term on the diffusivity of a jammed solid in Section \ref{stressed}. 

\begin{figure}
\includegraphics[width=0.45\textwidth]{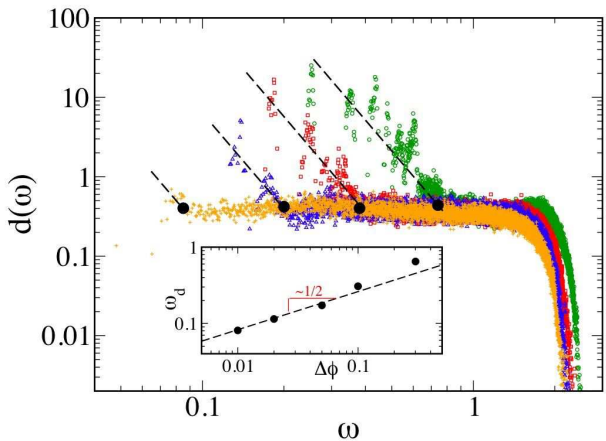}
\caption{\label{figharmonic} (Color Online) Plots of diffusivity, measured in units of $\sigma \sqrt{\epsilon/M}$ versus frequency, measured in units of $ \sqrt{\epsilon/\sigma^2 M}$  for an unstressed packing of 2000 particles with harmonic interactions at packing fractions $\Delta \phi=0.3$ (green), 0.1 (red), 0.05 (blue), 0.01 (yellow).  The black dots indicate the crossover frequency $\omega_{d}$ at each $\Delta \phi$, while the dashed lines show a power-law of $\omega^{-4}$, expected for weakly scattered plane waves.  The plateau diffusivity is $d_0 \approx 0.35$. The inset shows the packing fraction scaling of  $\omega_{d}$.}
\end{figure}

\section{\label{results} Energy transport crossover in model jammed solids}

Figure \ref{figharmonic} shows a scatter plot of the mode diffusivity $d(\omega_{i})$ obtained from evaluating numerically Eq.(\ref{eq:kubo5}-\ref{eq:Sij}) at four packing fractions $\Delta \phi=0.3$ (green), 0.1 (red), 0.05 (blue), 0.02 (yellow) for a 2000-particle packing. Three distinct transport regimes can be clearly identified in each curve, corresponding to ballistic, diffusive and localized vibrational modes. The crossover frequencies that mark the ballistic to diffusive crossover for each $\Delta \phi$, indicated in Fig. \ref{figharmonic} as black dots, do not depend on $N$ for systems of this size or larger. In what follows, the mode diffusivity data is presented as scatter plots of single particle configurations for clarity. We have tested that performing a frequency binning followed by a disorder-average over several distinct configurations confirms our conclusions \cite{Xu09}.  

At very high frequencies, Anderson localization sets in and the diffusivity drops rapidly as a result of localization of the 
vibrational modes. The contribution of localized modes to the thermal conductivity is negligible if anharmonic effects such as hopping are ignored.
Fig.~\ref{figharmonic} shows that the localization frequency for particles interacting with a repulsive harmonic potential does not depend strongly $\Delta \phi$ close to the jamming point. 

At low frequencies, the energy diffusivity exhibits a strong frequency dependence characteristic of vibrational modes that are essentially phonons weakly scattered by disorder. As a comparison we have drawn black dashed lines in Fig. \ref{figharmonic} indicating the power law divergence with $\omega^{-4}$ expected for Rayleigh scattering of plane waves incident on uncorrelated scattering centers.  Close inspection of the scatter plot reveals that the low $\omega$ peaks occur close to the discrete frequencies allowed in our cubic simulation box of size $L$ by the linear dispersion: 
\begin{equation}
\omega_{i} = \frac{2 \pi v}{L} \sqrt{p^2+q^2+r^2}
\end{equation}    
where $\{p,q,r\}$ denote the quantum numbers for the periodic system and the speed of sound $v$ is the transverse one for most low $\omega$ modes near $\phi_c$, see Eq. (\ref{eq:scaling2bis}-\ref{eq:scaling2}). In the continuum limit we expect that their density of states at very low $\omega$ will be given by the Deybe law $D(\omega)\sim \frac{\omega^2}{v^3}$. 

The intermediate frequency regime is the one of most direct relevance to the intermediate temperature behavior of the thermal conductivity.  Strikingly, this regime is characterized by a diffusivity, henceforth labeled as $d_{0}$, that is nearly independent of frequency (Fig. \ref{figharmonic}). The notion of a frequency-dependent diffusivity  has a long history dating back to Kittel's observation \cite{Kittel} that the experimental curve for $\kappa(T)$ in many glasses at room temperature could be interpreted in terms of a nearly frequency-independent mean free path of the order of a molecular length.  

The onset of the plateau in the diffusivity is marked by a crossover frequency, that exhibits a peculiar scaling with the packing fraction $\Delta \phi$. We henceforth label it as $\omega_{d}$.  The inset of Fig.~\ref{figharmonic} reveals that $\omega_{d} \sim \Delta \phi ^{0.5}$ for a system composed of harmonic springs. This is the same scaling with $\Delta \phi$ observed for the frequency $\omega^{*}$ above which a large excess of vibrational modes have been observed in previous studies of the density of states (see Eq. (\ref{eq:scaling3}) with $\alpha=2$). 

\subsection{\label{scaling} Dimensional analysis}
 
Fig.~\ref{figharmonic} shows that $d(\omega)$ is characterized by a well-defined crossover from ballistic to diffusive behavior. Our aim in this section is not to provide a rigorous 
derivation of the functional dependence of the diffusivity on frequency, but rather to understand how the defining features of the diffusivity (the height of the plateau, $d_0$, and the scaling of the crossover frequency, $\omega_{d}$) depend on applied pressure. This is done by keeping track of how the fundamental parameters that enter the definition of $d(\omega)$ scale with the packing fraction $\Delta \phi$.    First, however, we must understand how these parameters depend on the dimensional parameters in our system: the particle diameter, $\sigma$, the particle mass, $m$, and the energy scale for the potential, $\epsilon$.

The diffusivity has dimensions of $\frac{[Length]^2}{time}$. The natural unit of frequency in a vibrational system, by which the $\omega$ axis in Fig. \ref{figharmonic} is measured, is $\sqrt{\frac{k}{M}}$ where $k$ is the $bare$ elastic coupling of the solid.   More generally an effective spring constant, $k_{eff}$, can be defined by differentiating twice the interaction energy $V(r_{ij})$ of Eq. (\ref{eq:C}) and evaluating the result at the average equilibrium bond length  $<r^{eq}_{ij}>$ of interacting neighbors:
 
\begin{equation}
k_{eff} = \left . \frac{\partial^2 V(r_{nm})}{\partial r_{nm}^2} \right |_{r_{nm}=<r_{nm}^{eq}>} \!\!\!\!\! \quad . \label{keffdef}
\end{equation}

There is a simple linear relation between the relative change of $<r^{eq}_{nm}>$ upon compression and the corresponding macroscopic change of volume $\Delta \phi$ 
\begin{equation}
\frac{\sigma - <r^{eq}_{nm}>}{\sigma} \approx \frac{\Delta \phi}{3}  \!\!\!\!\! \quad , \label{deltar}
\end{equation}
where the numerical prefactor in the right hand side of Eq. \ref{deltar} was checked numerically \footnote{It corresponds to one over the number of 
dimensions if the deformations induced in the packing upon compression are affine.}. 

From Eqs.~(\ref{keffdef}-\ref{deltar}) we see that  
\begin{equation}
k_{eff} \approx \epsilon \frac{\alpha -1}{\sigma^2} \left(\frac{\Delta \phi}{3}\right)^{\alpha-2}  \!\!\!\!\! \quad ,
\label{trivial1}
\end{equation}
For harmonic repulsions, $\alpha=2$ and $k_{eff}=\epsilon/\sigma^2$ is independent of $\Delta \phi$, while for Hertzian potentials, $k_{eff} \sim \Delta \phi^ {1/2}$.

Note that the bulk modulus obeys the same scaling \cite{Durian95,O'Hern03}, so
\begin{equation}
B \sim k_{eff}  \!\!\!\!\! \quad .
\label{trivial2}
\end{equation}

The dimensional length scale in this problem is the particle size, $\sigma$. 
As a result, the diffusivity is naturally measured in units of the product of $\sigma^2$ times the characteristic frequency scale 
\begin{equation}
d(\omega) \sim \sigma^2 \sqrt{\frac{k_{eff}}{M}} \!\!\!\!\! \quad .
\label{d}
\end{equation}
Inspection of Fig. \ref{figharmonic} reveals that, as $\omega$ increases, the diffusivity decreases rapidly until it saturates at a value that we denote by $d_{0}$.  
Note from Fig. \ref{figharmonic} that $d_0 \approx 0.35 \sigma^2 \sqrt{k_{eff}/M}= 0.35 \sigma \sqrt{\epsilon/M}$, see Eq. (\ref{d}).  

\begin{figure}
\includegraphics[width=0.45\textwidth]{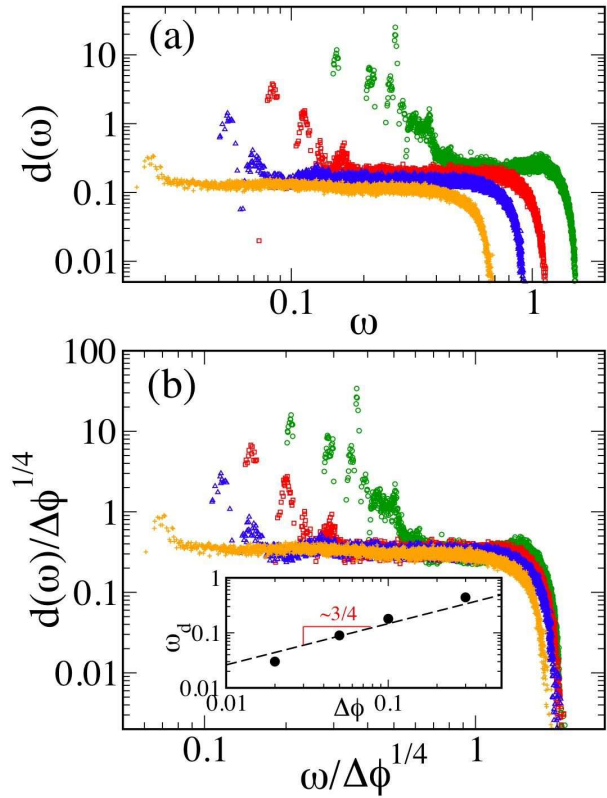}
\caption{\label{figherzian2} (Color Online) Plots of diffusivity vs.~ frequency for an unstressed packing of 2000 particles interacting via a Hertzian potential. The packing fractions are approximately $\Delta \phi=0.3$ (green), 0.1 (red), 0.05 (blue), 0.02 (yellow).  (a) $d(\omega)$ in units of $\sigma \sqrt{\frac{\epsilon}{M}}$) vs.~$\omega$ (in units of $ \sqrt{\frac{\epsilon}{\sigma^2 M}}$) (b) Scaled diffusivity, $d(\omega)/\Delta \phi^{1/4}$, vs. scaled frequency, $\omega/\Delta \phi^{1/4}$, showing data collapse in the plateau region.  The inset shows the packing fraction scaling of the crossover frequency $\omega_{d}$.
}
\end{figure}

Similarly, upon evaluating Eqs. \ref{trivial1} and \ref{d} for a Hertzian potential ($\alpha=5/2$) in three dimensions, we obtain the following prediction for the value of $d_{0}$
\begin{equation}
d_{0}= c  \frac{3^{1/4}}{2^{1/2}} \sigma \sqrt{\frac{\epsilon}{M}} \Delta \phi^{1/4}\quad \text{Hertzian} \!\!\!\!\! \quad .
\label{dh}
\end{equation}
$d_0 = c \sigma \sqrt{\epsilon/M} \Delta \phi^{1/4}$ where $c$ is of order unity.  Fig.~\ref{figherzian2}(a)  shows $d(\omega)$ for a Hertzian system at four different packing fractions.  Clearly, $d_0$ increases with $\Delta \phi$.  If we divide $d(\omega)$ and $\omega$ by $\Delta \phi^{1/4}$ as in Fig.~\ref{figherzian2}(b) we get a good collapse of the data, indicating that $d_0$ obeys the prediction, with $c=0.35$. The numerical value of $c$ is independent of the potential used and will be derived purely from the random geometry of a marginally jammed packing in Section \ref{micro}. 

These results for harmonic and Hertzian potentials suggest that the plateau in the diffusivity is consistent with what is loosely referred to as the "minimal conductivity hypothesis." \cite{Kittel,Slack}. The diffusivity $d_{0}$ is minimal in the sense that the length scale that multiplies the characteristic frequency is the smallest length that can be chosen in the system, the particle diameter $\sigma$.  Once $d(\omega)$ attains this minimal value it cannot decrease by much as $\omega$ increases, unless a transition occurs to a new transport regime characterized by a vanishingly small diffusivity. This is precisely what happens at the end of the plateau where Anderson localization sets in.  

\subsection{\label{test} Scaling of transport crossover}

The scaling of $\omega^*$ has been derived for systems near the isostatic jamming transition using a variational argument 
that predicts  the presence of extended heterogeneous modes with strong spatial de-correlations \cite{matthieu}. This structural property suggests that the ability to transport heat for vibrational modes above $\omega^*$ should be impaired \cite{matthieu2}.
In this section, we provide a heuristic argument which also suggests that  $\omega_{d} \sim \omega^{*}$ independent of the repulsive potential (e.g., for all $\alpha$). 

Inspection of Fig. \ref{figharmonic} and Fig. \ref{figherzian2} shows that for the range of packing fraction accessible to our simulation, the diffusivity appear to be smooth across the 
frequency $\omega_d$. At the crossover, where the diffusivity plateau $d_{0}$ begins and the ballistic regime has just ended, the kinetic formula for the diffusivity, Eq.~\ref{eq:Cc} leads 
\begin{equation}
d_{0}=\frac{1}{3} v_{t} \!\!\!\! \quad \ell_{d}  \!\!\! \quad ,
\label{eq:do}
\end{equation}
where we have assumed that an effective transverse speed of sound can be assigned to the vibrational modes up to the crossover frequency $\omega_d$ and that it exhibits a smooth crossover above it. This working assumption is corroborated in Section \ref{sec:disp} where the dispersion relation is determined numerically. The phonon density of states at low $\omega$ is dominated by shear waves near Point J as apparent from Equations (\ref{eq:scaling2bis}-\ref{eq:scaling2}) and the discussion following it.

We can now solve Eq. (\ref{eq:do}) for the mean free path $\ell_{d}$ at the crossover and compare it to wavelength $1/q_{d}$, which we can easily extract from the dispersion relation
$\omega_{d}=v_t q_{d}$. We find that over our range of compression
\begin{equation}
q_{d} \!\!\! \quad  \ell_{d} \sim 2 \!\!\! \quad .
\label{eq:ioffe}
\end{equation} 
The crossover frequency, $\omega_{d}$, corresponds to the "largest" wave-vector $q_{d}$ that can be meaningfully defined in our disordered packings and is consistent with the Ioffe Regel criterion \cite{Ioffe}. That is to say, the wavelength $1/q_{d}$ is of order of the mean free path $\ell_{d}$ \cite{Ioffe} over the range of compression probed in our study. The main requirement to extend the validity of the Ioffe Regel criterion for smaller values of $\Delta \phi$, is that the number on the left-hand side of Eq. (\ref{eq:ioffe}) remains weakly dependent on packing fraction. 

Upon substituting the mean free path $\ell_{d}$ from Eq. (\ref{eq:ioffe}) into Eq. (\ref{eq:do}), we obtain
\begin{equation}
1/q_{d} \sim \frac{d_0}{v_t} \sim \frac{\sqrt{k_{eff}}}{v_t} \!\!\!\! \quad ,
\label{eq:do2}
\end{equation}
where the overall scaling of the diffusivity curve indicated in Eq. (\ref{d}) was used to write the last step and $\ell_{d} \sim 1/q_{d}$.

Upon setting the transverse sound speed proportional to the square-root of the shear modulus $G$ we find 
\begin{equation}
q_{d} \sim \sqrt{\frac{G}{k_{eff}}}\sim \Delta \phi^{1/4} \!\!\!\! \quad ,
\label{eq:do3}
\end{equation}
where Eqs.~(\ref{eq:scaling0}-\ref{eq:scaling1}) were substituted for the elastic moduli. As $\phi \rightarrow \phi_c$, the $dynamic$ wavevector $q_{IR}$ vanishes with a power-law 
exponent of $1/4$, independent of the underlying inter-molecular potential.  This scaling is consistent with previous numerical studies of the peak position of the {\it transverse} structure factor at the frequency onset of excess modes, which found a scaling of $\Delta \phi ^{0.24 \pm 0.03}$ \cite{Leo1}.   
The non-trivial hypothesis that the diffusivity is smooth across $\omega_d$ implies that $\ell_{d} \sim \Delta \phi^{-1/4}$.

Upon substituting Eq. (\ref{eq:do2}) in the dispersion relation and using Eqs.~\ref{eq:scaling00}-\ref{eq:scaling0}, we obtain
\begin{equation}
\omega_{d} \sim \frac{v_t^2}{k_{eff}} \sim \frac{G}{\sqrt{k_eff}} \sim \Delta \phi ^{\frac{\alpha - 1}{2}}  \sim (z-z_c)^{\alpha-1}  \!\!\! \quad .
\label{eq:do3a}
\end{equation} 

It is straightforward to conclude from Eq. (\ref{eq:do3a}) that $\omega_{d}$ should scale as $\Delta\phi^{\frac{1}{2}}$ and $\Delta\phi^{\frac{3}{4}}$ for harmonic and hertzian interactions respectively. This conclusion is consistent with the numerical results plotted in the insets of  Figs.~ \ref{figharmonic} and \ref{figherzian2}(b). 

The limited dynamic range of the simulation stems from the fact that the crossover cannot be seen if the wavelength $1/q_{d}$ exceeds the system size. The disappearance of the low-$\omega$ diffusivity upturn at low $\Delta \phi$ is clearly seen in Figs.~\ref{figharmonic}-\ref{figherzian2}. In future work, we hope to probe in more detail the behavior of the diffusivity in the crossover region. 

\subsection{\label{sec:boson} Relation of transport crossover to boson peak}

In this section we show numerically that the transport crossover frequency $\omega_{d}$ observed in the diffusivity plots of Figs. ~\ref{figharmonic}-\ref{figherzian2}, corresponds to the same frequency scale, $\omega^*$, at which the onset 
of the excess vibrational modes is observed in the density of states. The signal advantage of jammed sphere packings over models studied previously is that one can verify this identification at different packing fractions and hence test not only that the two frequency scales are close in numerical value for a given $\phi$ but also that they scale in the same way as a function of compression.

\begin{figure}
\includegraphics[width=0.45\textwidth]{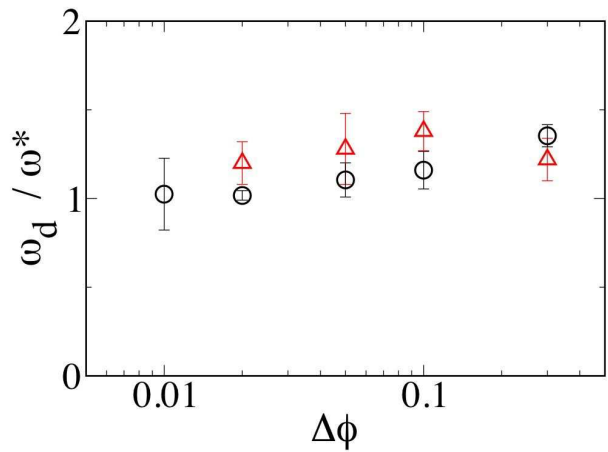}
\caption{\label{fig:boson} Plot of the transport crossover versus the boson peak at different $\Delta \phi$ in jammed packings of 2000 particles with harmonic (black open circles) and Hertzian (red open triangles) repulsions.  The ratio is near unity and is nearly constant with $\Delta \phi$.}
\end{figure}

We first note that the scaling for $\omega_{d}$ in Eq. (\ref{eq:do3a}) is identical with the earlier numerical observation of the boson peak frequency, $\omega^*$, in Eq. (\ref{eq:scaling3}) as well as with the relation derived theoretically in Ref. \cite{matthieu} for the frequency onset of the anomalous modes of compressed jammed packings.  Fig. \ref{fig:boson} shows the ratio $\omega_{d}/\omega^*$ as a function of compression, $\Delta \phi$.   Here, the frequency $\omega^{*}$ is measured numerically from the onset of the plateau in the density of states; see Ref. \cite{Xu07} for details.  The variation of the ratio $\omega_{d}/\omega^*$ is small compared with the variation of $\omega_{d}$, which changes by an order of magnitude over the same range of $\Delta \phi$ (see inset to Fig.~\ref{figharmonic}).    Thus, the transport crossover frequency and boson peak frequency track each other, implying that the same physics underlies both phenomena.  In particular, the result shows that the excess modes above the boson peak frequency have a small and nearly frequency-independent diffusivity.

\subsection{\label{sec:disp} Change in nature of modes at transport crossover}
 
The Fourier decomposition of the vibrational modes evolves dramatically as the frequency is increased through the transport crossover at $\omega_{d}$. We concentrate here on $f_{T}(q,\omega)$, the transverse Fourier components~\cite{Grest2,Leo2}:
\begin{eqnarray}
f_{T}(q,\omega) = \left\langle \left| \sum_{n} {\bf\hat{q}} \wedge
    {\bf e}_n(\omega) \exp(\imath {\bf q} \cdot {\bf r}_n)\right|^2
\right\rangle 
\label{eq:ft}
\end{eqnarray}
where $q$ denotes the wavevector and ${\bf e}_n(\omega)$ is the polarization vector of particle $n$ of the mode at frequency, $\omega$.  The brackets indicate an average over directions of $\hat q$.  

\begin{figure}
\includegraphics[width=0.45\textwidth]{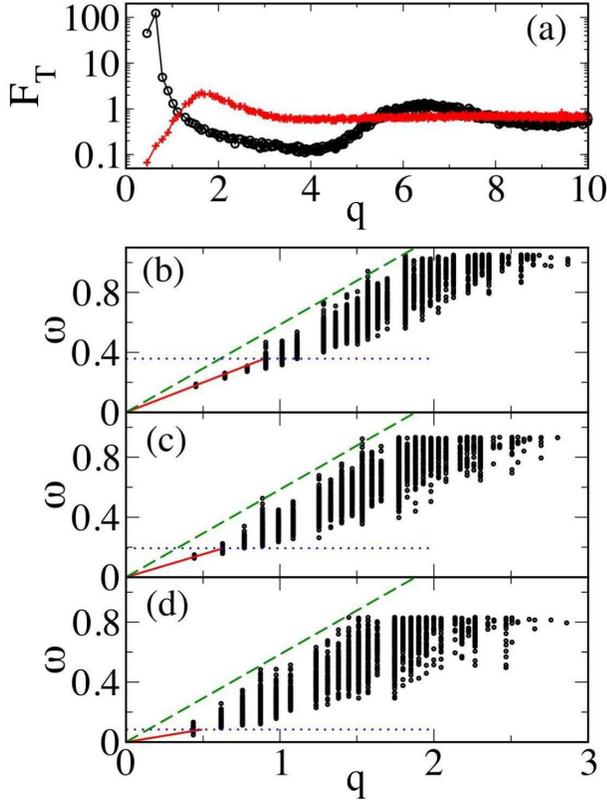}
\caption{\label{fig:dispersion} (Color online) Transverse mode structure factor $f_{T}(q,\omega)$ for $N=2000$ and $\Delta \phi=0.1$ at $\omega=0.23$ (black) below the transport crossover at $\omega_{IR}=0.36$ and at $\omega=0.74$ (red) well above the crossover.  (b-d) Phono n dispersion relations for (b) $\Delta\phi=0.1$, (c) $0.05$, and (d) $0.01$, respectively. The data are at discrete values of $q$ because of the finite system size. The horizontal (blue) dotted line in each panel (b-d) marks $\omega_{d}$, while the solid line (red) marks the transverse sound speed that varies with compression as given by Eq.~\ref{eq:scaling2bis}.  As a comparison, the dashed (green) line is the same in each panel (independent of compression) and has a slope proportional to the speed of longitudinal sound.  Note that the solid line gives the dispersion for frequencies below $\omega_{d}$.}
\end{figure}

At all compressions, $f_{T}(q,\omega)$ has a low-wavevector peak at $q=q_{max}$ that shifts to higher values with increasing frequency. Two typical examples are shown in Figure \ref{fig:dispersion}(a) at $\Delta \phi=0.1$. Below $\omega_{d}$ (black curve), where the diffusivity decreases rapidly with increasing frequency, the peak is sharp and tall.  Here the modes resemble weakly-scattered transverse plane waves with wave-vector $q_{max}$. By contrast above $\omega_{d}$ (red curve), in the region of the diffusivity plateau, the peak is dramatically less pronounced; the peak height is $\sim 50$ times smaller than for the black curve and is comparable to the background signal observed at higher $q$. This is the characteristic signature of the strong-scattering regime where vibrational modes can no longer be meaningfully characterized by a narrow range of $q$. Such modes are poor conductors of energy, as reflected in the low value of $d(\omega)$ above $\omega_{d}$.  This evolution in character is not sharp and the peak height decreases continuously as $\omega$ increases past $\omega_{d}$; a peak - albeit a small one - appears even for modes in the diffusivity plateau. 

From data of $q_{max}$ versus $\omega$ one can determine the transverse-sound dispersion curve~\cite{Grest2,Leo2}. Fig. \ref{fig:dispersion}(b-d) shows the transverse dispersion relation for three of the packing fractions shown in the diffusivity plot of Fig. \ref{figharmonic}. Each point represents the value of $q_{max}$ obtained from the peak in $f_{T} (q,\omega)$ for a single vibrational mode of frequency $\omega$. In other words, each point represents the wavevector that makes the largest contribution to a vibrational mode. At each packing fraction the crossover frequency $\omega_{d}$ is represented by a horizontal dotted line. The solid red line, $\omega=v_t q$, shows the expected transverse dispersion relation with $v_T$ decreasing with decreasing $\Delta \phi$ on the basis of Eq. (\ref{eq:scaling2bis}). The dashed green line in each panel has the same slope independent of $\Delta\phi$. From Eq. (\ref{eq:scaling2bis}), its slope is therefore proportional to (but smaller than) the longitudinal sound speed, $v_l$.
  
These dispersion curves also show a marked change in behavior as the frequency is varied through $\omega_{d}$.  Below the crossover frequency, the peaks are not only sharp, as indicated by the behavior in Figure \ref{fig:dispersion}(a), but their position corresponds very well with that given by the transverse speed of sound (red solid line). Above $\omega_{d}$, on the other hand, the peaks are squat and broad and no longer follow the line given by the transverse sound speed. Instead, the lowest values of $q_{max}$ shift to smaller $q$ and begin to track the green line which has a slope independent of compression and proportional to the longitudinal, not the transverse, speed of sound.  The spread in the positions of $q_{max}$ for frequencies $\omega>\omega_{d}$ indicate that the peaks are very broad in this region.  Similar results for $f_{T}(q,\omega)$ and the transverse dispersion relation above $\omega_{d}$ were earlier found for a model that included the stress terms in the energy~\cite{Leo2}. In that case, the dispersion relation above $\omega_{d}$, showed essentially no variation in the position of the peaks, $q_{max}$, despite an increase in $\Delta\phi$ by five orders of magnitude. (The data in that study is slightly different from the data reported here in that it showed averaged $f_{T} (q,\omega)$ over several nearby frequency modes in a bin instead of the peaks in individual modes. Both ways of treating the data show the same general features.)

At frequencies below $\omega_{d}$, the transport crossover is heralded by when the weakly scattered plane waves begin to scatter strongly.  When the modes are only weakly scattered, the dispersion curves are marked by a single sharp peak at $q=\omega/v_t$. As the scattering becomes stronger, other wave vectors begin to contribute to $f_{T}(q,\omega)$ so that the peaks become very broad and includes a very wide range of wave vectors.  This is qualitatively what we see in Figure \ref{fig:dispersion}(a) where the small peak in the red curve is hardly greater than the background value. 
 
\section{\label{sec:plateau} Plateau in the diffusivity}   

In the previous section, we have shown that $\omega_{d}$, where the diffusivity flattens out, scales in the same way as does the frequency associated with the excess modes of the boson peak.  We can thus consider the flat diffusivity as the signature of the excess vibrational modes that appears in their transport properties.  In the present section, we will focus on why the diffusivity is flat over an extended frequency range above $\omega_{d}$.  We will address this (A) by showing that from the form of the matrix elements, the diffusivity should be simply proportional to the density of the modes themselves, which is also flat in this region; and (B) by examining what properties of the modes are necessary for producing a flat diffusivity.  Our study suggest that $\omega_{d} \sim \omega^*$ vanishes at Point J.  At this point, the plateau extends over the entire frequency range up to the onset of localization.  This simplifies the analysis.  Thus, Point J is a natural place to gain insight into the origin of the constant diffusivity.

\begin{figure}
\includegraphics[width=0.4\textwidth]{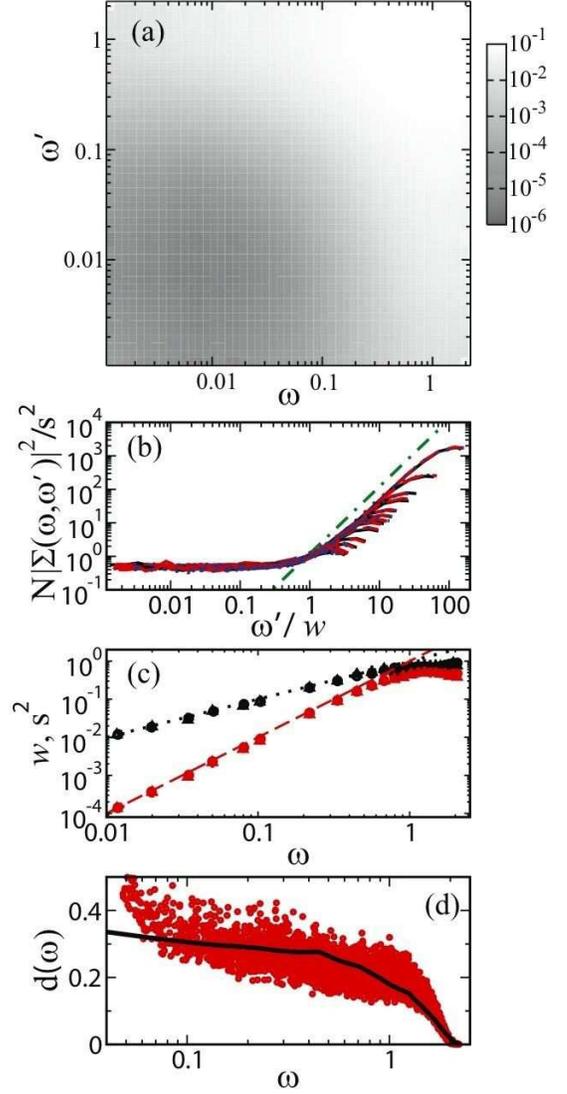}
\caption{\label{fig1} (Color Online) Diffusivity just above the jamming transition
at $\Delta \phi=10^{-6}$.  (a) Contour plot of the heat-flux matrix elements $|\Sigma(\omega, \omega' )|^2$
plotted versus $\omega$ and $\omega'$ at $N=2000$. (b) Scaling plot
showing collapse of $|\Sigma(\omega,\omega')|^2$ at $N=2000$ (black solid),
$1000$ (red dashed), and $500$ (blue dotted) with scale factors $s^2$ and
$w$. (c) Scale factors $s^2$ (red symbols) and $w$ (black symbols) versus $\omega$.  We find $s^2 \propto \omega^2$ (red dashed line)
and $w \propto \omega$ (black dotted line) except at high $\omega$. (d) Scatter Plot of $d(\omega)$ with $N=2000$ and delta function broadening $\eta=0.002$ (dashed line).  Solid line shows predicted $d(\omega)$ for the infinite system.}
\end{figure}

\subsection{Energy flux matrix elements}

We start by showing that the energy flux matrix elements have a particularly simple form at the jamming threshold, which enables us to determine the $N \rightarrow \infty$ behavior of the diffusivity~\cite{Xu09}. Consider the frequency averaged matrix elements defined as 
\begin{equation}
|\vec \Sigma(\omega,\omega^\prime)|^2=\sum_{ij} |\vec S_{ij}|^2 \delta (\omega-\omega_i)\delta(\omega^\prime-\omega_j)
\label{eq:Sigmadef}
\end{equation}
where $i$ and $j$ are indexes labeling the vibrational modes and the matrix elements $\vec{S}_{ij}$ are given by Eq. (\ref{eq:Sij}) in the limit $\omega_i \rightarrow \omega_j$.
 
Figure \ref{fig1}(a) shows the heat-flux matrix elements $|\vec \Sigma(\omega,\omega^\prime)|^2$ defined in Eq.(\ref{eq:Sigmadef}) for packings at $\phi-\phi_c=10^{-6}$ for 
different values of $\omega$ versus $\omega^\prime$.   Note that the matrix elements are symmetric in $\omega$ and $\omega^\prime$ and that they increase with increasing $\omega$ and $\omega^\prime$.  Fig. \ref{fig1}(b) shows that all the curves for different system sizes, $N$, and frequencies, $\omega$, can be collapsed onto a simple scaling form given by
\begin{eqnarray}
|\vec \Sigma(\omega,\omega^\prime)|^2 &\sim& \frac{1}{N}  \quad \!\!\! \omega'^2  \quad \quad \quad \text{if} \quad \omega'>\omega \nonumber \\
|\vec \Sigma(\omega,\omega^\prime)|^2 &\sim& \frac{1}{N}  \quad \!\!\! \omega^2  \quad \quad \quad \text{if} \quad \omega'<\omega
\label{eq:scalesigma}
\end{eqnarray}
except at high frequency where the modes
become localized~\cite{Grest,Leo2,zorana} and the curves in Fig. \ref{fig1}(b) start peeling off from the green line.  Figure \ref{fig1}(c) shows that the scale factors for the collapse satisfy $s^2=\omega^2$ and $w=\omega$, respectively, consistently with Eq. (\ref{eq:scalesigma}). The scaling collapse demonstrates that the only noticeable system-size dependence is a prefactor of $1/N$. Since for large $N$, the density of states scales as $N$ ~\cite{Leo1}, Eq. (\ref{eq:conductivity}) therefore yields a well-defined diffusivity in the $N \rightarrow \infty$ limit, shown as the solid curve in Fig.~\ref{fig1}(d).

The scaling collapse in Fig.~\ref{fig1}(b) implies that $|\vec \Sigma(\omega,\omega)|^2\propto\omega^2/N$ at low frequencies.   This result, combined with Eq.~\ref{eq:kubo5}, implies that
$d(\omega)\propto D(\omega)$ at low $\omega$.  This $N \rightarrow \infty$ prediction for $d(\omega)$ is shown as the solid line in Fig.~\ref{fig1}(c).   Thus, the diffusivity is nearly constant down to $\omega=0$ at Point J because the density of states is nearly constant there~\cite{Leo1}. 

Over most of the frequency range, this $N \rightarrow \infty$ prediction agrees very well with the scattered points in Fig.~\ref{fig1}(c), which show $d(\omega)$ for a system with N=2000.  At low frequency $d(\omega)$ deviates from the solid line and exhibits an upturn. This upturn is a finite-size artifact that arises from replacing the delta-function with a smoothing function in Eq. (\ref{eq:delta}).  It scales as $\omega^{-3}$ with a prefactor that vanishes as $N \rightarrow \infty$, $\eta \rightarrow 0$. 

\subsection{\label{micro} Properties of modes in the plateau}

It is important to understand what specific properties of the modes give rise to the plateau in the diffusivity.   To simplify the analysis, we will consider only systems of monodisperse particles interacting via harmonic repulsions ($\alpha=2$) just above the jamming threshold.  The argument can be generalized to the bidisperse case studied in this paper.

The starting point for deriving the flat diffusivity and for deriving its plateau value, $d_0$, is Eq. \ref{eq:Sij} for the matrix element of the heat flux operator evaluated using periodic boundary conditions. Recall that for two modes of frequencies within a bin centered at $\omega$ the matrix element $S_{ij}$ reads
\be
\label{3}
S_{ij}=\frac{hi}{2Vm} \sum_{n,m}  {\vec e}(n,i) H(x,y)  {\vec e}(m,j) ({\vec R_m}-{\vec R_n})
\ee
where $n$ and $m$ label the particles, $\vec R_n$ and $\vec R_m$ their positions, $H(n,m)$ is the dynamical matrix element (itself a $d*d$ matrix)  between these two particles, ${\vec e}(n,i)$ is the displacement of particle $n$ in mode $i$ and $M$ is the particle mass.  

Next set $ ({\vec R_m}-{\vec R_n})= \sigma {\hat R}_{nm}$, where $\sigma$ is the particle diameter and ${\hat R}_{nm}$ the unit vector from $n$ to $m$.  Set the non-diagonal terms $ H(n,m)= k {\hat R}_{nm}\otimes {\hat R}_{nm}$, where $k$ is the contact stiffness. With this substitution, Eq. (\ref{3}) can be rewritten as a sum on all contacts $\langle n,m\rangle$:
\begin{eqnarray}
\label{4}
S_{ij}&=&\frac{ih\sigma k}{2V M} \sum_{\langle n,m\rangle}  {\hat R}_{nm} [ ( {\vec e}(n,i)\cdot {\hat R}_{nm}) (  {\vec e}(m,j) \cdot {\hat R}_{nm}) \nonumber \\
&-&({\vec e}(n,j)\cdot {\hat R}_{nm}) (  {\vec e}(m,i) \cdot {\hat R}_{nm})]
\end{eqnarray}
We can then rewrite the sum in Eq.(\ref{4}) as:
\begin{eqnarray}
\Sigma(\omega)&=& \sum _{\langle n,m\rangle}  {\hat R}_{nm} [ ( {\vec e}(n,i)\cdot {\hat R}_{nm} - {\vec e}(m,i) \cdot {\hat R}_{nm})({\vec e}(m,j)\cdot {\hat R}_{nm}) \nonumber \\
&-&( {\vec e}(n,j)\cdot {\hat R}_{nm} - {\vec e}(m,j) \cdot {\hat R}_{nm})({\vec e}(m,i)\cdot {\hat R}_{nm}) ] 
\end{eqnarray}  
and 
\begin{eqnarray}
\Sigma(\omega)&=& \sum _{\langle n,m\rangle}  {\hat R}_{nm} [(\delta r_{nm}^i)({\vec e}(m,j)\cdot {\hat R}_{nm}) \nonumber \\
&-&( \delta r_{nm}^j)({\vec e}(m,i)\cdot {\hat R}_{nm}) ]
\end{eqnarray}
where $\delta r_{nm}^i\equiv ( {\vec e}(n,i)\cdot {\hat R}_{nm} - {\vec e}(m,i) \cdot {\hat R}_{nm})$ is the stretching of the contact $nm$ corresponding to mode $i$. Taking the amplitude squared of $\Sigma$ leads to diagonal and non-diagonal terms. The latter are of two forms:
\ba
 (\delta r_{nm}^i {\hat R}_{nm} )(  \delta r_{pq}^i {\hat R}_{pq}) ({\vec e}(m,j)\cdot {\hat R}_{nm}) ({\vec e}(q,j)\cdot {\hat R}_{pq}) \\
( \delta r_{nm}^i {\hat R}_{nm})(   \delta r_{pq}^j {\hat R}_{pq}) ({\vec e}(m,j)\cdot {\hat R}_{nm}) ({\vec e}(p,i)\cdot {\hat R}_{pq}) 
 \ea
 where $nm$ and $pq$ correspond to two distinct contacts. 
 
We now make some assumptions about the nature of the modes whose validity we test numerically. (i) if the displacements are uncorrelated between different modes ($ \langle {\vec e}(m,j)\cdot {\vec e}(q,i) \rangle=0$), the two terms above become:
 \ba
 \label{zoo}
\langle (\delta r_{nm}^i {\hat R}_{nm} )(  \delta r_{pq}^i {\vec n}_{pq})\rangle\langle ({\vec e}(m,j)\cdot {\hat R}_{nm}) ({\vec e}(q,j)\cdot {\hat R}_{pq}) \rangle\\
\langle ( \delta r_{nm}^i {\hat R}_{nm})({\vec e}(q,i)\cdot {\hat R}_{pq})  \rangle \langle({\vec e}(m,j)\cdot {\hat R}_{nm})(   \delta r_{pq}^j {\hat R}_{pq})\rangle
 \ea
This assumption was tested numerically in a packing comprised of $N=2000$ particles at $\Delta \phi= 10^{-6}$ for the low frequency modes. Each of the two terms was found to be of the order of $10^{-21}$ which is vanishingly small within numerical precision. 
 
 We next assume that modes have weak spatial correlations.  Numerical simulations actually support  that such correlations exist \cite{Leo2} and grow as the frequency decreases. Neglecting them nevertheless appears to yield good quantitative results, as we shall see below. In particular,  each of these two terms in Eq.(\ref{zoo}) vanish under the specific assumptions that (ii) the directions of the stretching of different contacts within a mode are not correlated in space ($\langle (\delta r_{nm}^i {\hat R}_{nm})(\delta r_{pq}^i {\hat R}_{pq})\rangle=0$) and that (iii) the directions of stretching and of displacement are locally uncorrelated ($\langle (\delta r_{nm}^i {\hat R}_{nm})({\vec e}(q,i)\cdot {\hat R}_{pq})\rangle=0$). For assumption (ii) we find that the corresponding terms are of the order of $10^{-11}$ or
lower. For assumption (iii) we find a term of the order of
$10^{-13}$ or less.  If the quantities were correlated, we should find that for extended modes they are of the order of $1/N \approx 10^{-4}$, which is much larger than the values shown above. Therefore, the assumptions listed above appear to be reasonable and we are left with:
\be
|\Sigma|^2\approx  2 \sum _{\langle n,m\rangle}   \delta r_{nm}^i{}^2 ({\vec e}(m,j)\cdot {\hat R}_{nm})^2\\
\ee
Assuming now that (iv)  the amplitude of the displacements between two modes are not correlated (which is wrong for localized modes, where the displacements are anti-correlated since localized modes do not live on the same regions of space) then we have:
\be
|\Sigma|^2 =  2 \sum _{\langle n,m\rangle}  \langle \delta r_{nm}^i{}^2 \rangle \langle ({\vec e}(m,j)\cdot {\hat R}_{nm})^2 \rangle\\
\ee
The modes are normalized so $\langle ({\vec e}(m,j)\cdot {\hat R}_{nm})^2 \rangle=1/N{\cal D}$, where ${\cal D}$ is the dimensionality of space.  Also, the mode energy is $\delta E= M \omega^2/2\equiv k/2 \sum_{\langle n,m\rangle}  \delta r_{nm}^i{}^2$, so we obtain
\be
\Sigma=   \frac{2M\omega^2}{k} \frac{1}{N{\cal D}}
\ee
Then we have for Eq.(\ref{4}):
\be
|S|^2=\frac{h^2\omega^2\sigma^2k}{2V^2{\cal D}N}
\ee

This leads to:
\be
d(\omega)= \frac{\pi  D(\omega) \sigma^2 k}{3 m {\cal D}}
\ee
where $D(\omega)$ is the density of states per particle.  In units where $\sigma=k=m=1$, we obtain
\be
d_{0}= \frac{\pi D_0}{9}
\ee
in three dimensions, where $d_{0}$ and $D_{0}$ denote the plateau values of the diffusivity and the density of states, respectively. This is consistent with the numerical data in Fig. \ref{fig1}(d), which shows that the diffusivity
roughly follows the nearly flat density of states at Point J.

In summary, we obtain a frequency-independent diffusivity when the density of states is frequency-independent and the following conditions are satisfied:	(A)  displacements of particles in different modes of similar frequency are uncorrelated; (B)  the directions of  changes in the relative displacements of pairs of interacting particles are spatially uncorrelated within a given mode; and (C) the direction of change in the relative displacement of a pair of interacting particles is uncorrelated from the direction of the displacement.

\subsection{\label{stressed} Stressed packings}

Until now, we have neglected the forces that particles within a jammed packing exert on each other by replacing compressed springs between particles with springs at their equilibrium length (see Sec. \ref{unstressed}).  Here we restore the stress terms into the dynamical matrix and examine how they affect the behavior.  

\begin{figure}
\includegraphics[width=0.45\textwidth]{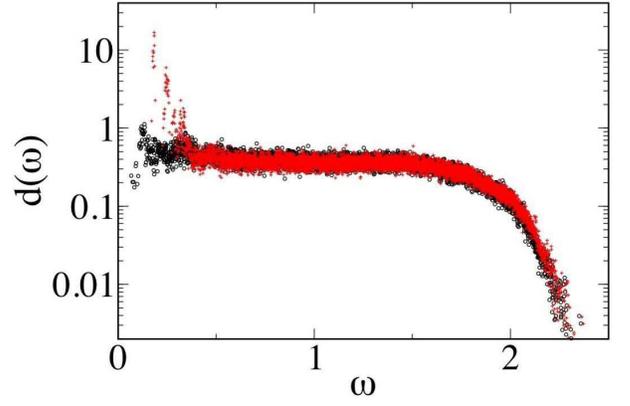}
\caption{\label{comparison} (Color Online) Diffusivity for $\Delta \phi=0.1$ in stressed (black) and unstressed (red) packings composed of $1000$ particles with harmonic repulsions.}
\end{figure}

The repulsive forces between particles tend to destabilize the system towards buckling, or rearrangements.  As a result, they tend to push modes to lower $\omega$. As a result, finite-size effects, which cut off plane waves at low frequency, are more obstructive.  This prevents us from studying the transport crossover directly in stressed systems.

\begin{figure}
\includegraphics[width=0.45\textwidth]{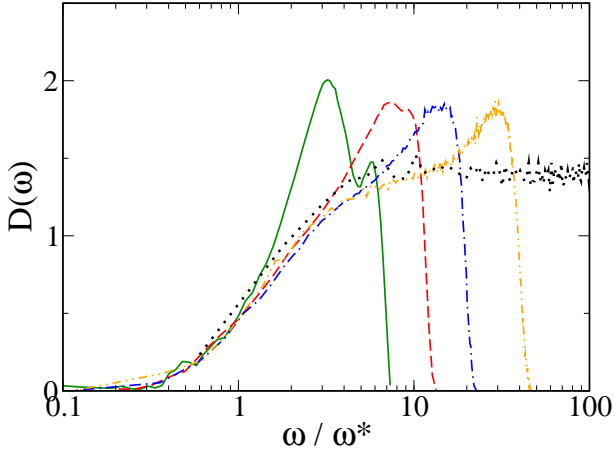}
\caption{\label{dosstressed} (Color Online) Density of states $D(\omega)$ vs. $\omega/\omega^*$ for $\Delta \phi=0.0001$ (black dotted), $0.01$ (yellow dashed-dot-dot), $0.05$ (blue dot-dashed), $0.1$ (red dashed) and $0.3$ (green solid), for a stressed system with $N=2000$  with harmonic repulsions.  Note that $\omega^*$ scales according to Eq.~\ref{eq:scaling3}.  This scaling produces a good collapse of the vibrational spectrum at low frequency.}
\end{figure}

Nevertheless, we suggest that the scaling of $\omega_{d}$ with compression should be the same in the stressed case as in the unstressed one.  This follows from the fact the boson peak frequency, $\omega^*$, follows the same power law (Eq.~\ref{eq:scaling3}) as in the unstressed case.  This is illustrated in Fig.~\ref{dosstressed}, which shows that the low-frequency portion of the vibrational spectrum collapses onto a single curve when $\omega$ is scaled by $\omega^* \sim\Delta \phi^{1/2}$ for harmonic repulsions.  
  
\section{ \label{ac} AC thermal conductivity at point J.}

The calculation of the energy flux matrix elements in Eq. (\ref{eq:scalesigma}) opens up the possibility of estimating the thermal conductivity $\kappa(T,\Omega)$ of the marginally jammed solid in the presence of an AC thermal gradient driven with angular frequency $\Omega$. As a starting point we rewrite Eq. (\ref{eq:kubo3}) as an integral over $\omega$
rather than a double sum over eigenmodes 
\begin{eqnarray}
\kappa(T,\Omega) &=&-\frac{\pi V}{T \Omega} \int \left [ n(\omega +\Omega)-n(\omega)\right] \left | \left<\omega+\Omega\left|S\right|\omega\right> \right|^2 \times \nonumber \\ &\times& D(\omega) \quad \!\!\!\! D(\omega+\Omega)  \quad \!\!\! d \omega \quad .
\label{eq:kappacont}
\end{eqnarray}
where the frequency averaged matrix elements read
\begin{eqnarray}
\left | \left<\omega+\Omega\left|S\right|\omega\right> \right|^2 &=& \frac{(2 \omega +\Omega)^2}{4 \omega  (\omega +\Omega)} |\Sigma(\Omega +\omega,\omega) |^2  \label{eq:e1}\\ 
&\approx& \frac{\omega^2}{N} \left(1+\frac{2\Omega}{\omega} \right) \quad .
\label{eq:matS}
\end{eqnarray}
and higher order corrections in terms of $\Omega^2$ have been dropped. The prefactor in Eq. (\ref{eq:Sij}), which was set to unity in evaluating $\Sigma(\omega,\omega')$, has been explicitly included since the mode coupling between $\omega$ and $\omega'$ is no longer restricted to vibrational states at the same frequency. 
Despite the concise notation adopted, Eq. (\ref{eq:kappacont}) accounts for both upwards and downwards jumps, $\omega \rightarrow \omega \pm \Omega $ , corresponding to energy being absorbed or injected into the reservoirs \cite{Mott}. In order to obtain Eq. (\ref{eq:matS}), Eq.  (\ref{eq:scalesigma}) was substituted into Eq. (\ref{eq:e1}).

With the aid of Eq. (\ref{eq:clong}), we can expand the difference in occupation numbers to first order in $\Omega$  
\begin{eqnarray}
\frac{n(\omega +\Omega)-n(\omega)}{\Omega} &\sim& - \frac{T}{\hbar \omega^2} \quad \!\!\!\! C(\omega) \nonumber \\
&-& \Omega \quad \!\!\!\! \frac{\partial}{\partial \omega} \left[ \frac{T}{\hbar \omega^2} \quad \!\!\!\! C(\omega)\right] \quad \!\!\! .
\label{eq:nN}
\end{eqnarray}
Upon substituting Equations (\ref{eq:nN}) and (\ref{eq:matS}) into Eq. (\ref{eq:kappacont}), four terms are obtained of which one is $ \text{\cal{O}}(\Omega^2)$ and it will be ignored.
After performing an integration by parts on the $\text{\cal{O}}(\Omega)$ term involving the $\omega$ derivative and canceling out two terms which are equal and opposite, we obtain
\begin{eqnarray}
\frac{\hbar \quad \!\!\!\! \kappa (T,\Omega)}{\pi c_{0} N_{0}^2} \approx  \int_{0}^{\omega_{max}} \! C(\omega) \quad \!\!\!\!  d\omega +  \Omega \left[C(\omega) \right]^{\omega_{max}}_{0} \quad \!\!\!\! ,
\label{thermocon}
\end{eqnarray}
where $N_{0}$ is the value of the plateau in the density of states. 

The desired result follows, according to Eq. (\ref{eq:clong}), upon setting $C(\omega_{max}=\infty)\approx 0$ 
\begin{eqnarray}
\kappa (T,\Omega) \approx  \frac{\pi c_{0} N_{0}^2 k_{B}^2}{\hbar^2} \left( \alpha \quad \!\!\!\! T  - \frac{\hbar \Omega}{k_B} \right) \quad \!\!\!\!\! +  \quad \!\!\!\!\! \text{\cal{O}}(\Omega^2) \quad \!\!\! ,
\label{thermoconb}
\end{eqnarray}
where the numerical constant $\alpha$ is given by
\begin{eqnarray}
\alpha \approx \int_{0}^{\infty} \frac{x^2 e^{x}}{(e^x -1)^2} \quad \!\!\!\! dx \quad \!\!\! .
\label{eqn}
\end{eqnarray}
As expected intuitively, a non-vanishing driving frequency $\Omega$ results in a lower thermal conductivity. 

Eq. \ref{thermoconb} allows the calculation of the $T$-dependent DC thermal conductivity ($\Omega \rightarrow 0$) at point $J$.  We note that this is particularly simple and can be calculated at the harmonic level without facing any divergences because no acoustic phonons are present.  From Eq. \ref{thermoconb}, we find that $\kappa(T)$ grows linearly in temperature for small $T$ and saturates above a temperature $k_B T_{max}\approx \hbar \omega_{max}$, where $\omega_{max}$ is the maximum frequency above which there are no more vibrational states. This follows from assuming that both the density of states and the diffusivity are approximately $\omega$ independent as expected from the scaling analysis and numerical extrapolations presented in Section \ref{scaling}.   Thus, a plateau in the diffusivity leads to an approximately linear increase of $\kappa(T)$ followed by a saturation.

\section{ \label{conclusion} Conclusion}

We have studied a class of model amorphous solids whose elastic properties can be tuned by varying the density near the jamming/unjamming transition, Point J.  The proximity to Point J allows variation of the crossover frequency that marks the onset of the plateau in the diffusivity, $\omega_{d}$, with $\Delta \phi$.  As $\Delta \phi \rightarrow 0$, our scaling arguments show that $\omega_{d} \rightarrow 0$, so that the plateau in the diffusivity extends all the way down to zero frequency.  Moreover, the value of $\omega_{d}$ agrees with the boson peak frequency $\omega^*$ within a factor close to unity; they both scale the same way with $\Delta \phi$.    

The findings presented in this study enable us to establish that (1) there is a frequency regime in which the diffusivity is small and nearly constant, and (2) the boson peak frequency coincides with the energy transport crossover frequency for all pressures applied to our unstressed amorphous packings of repulsive spheres. More work is needed to assess the relationship between the boson peak and the transport cross-over when pre-stress is important. Nonetheless, our results suggest solutions to two longstanding conundrums posed by these amorphous solids.   First, in such systems, the plateau in the thermal conductivity at intermediate temperatures is followed by a rise and then a saturation at high temperatures.  Crystals show the opposite behavior, with a thermal conductivity that decreases with $T$ at high $T$, see Fig. 5.1 in Ref. \cite{Phil81}.  What is the origin of the generic monotonicity of $\kappa$ with $T$ in amorphous solids?  Second, the temperature range of the plateau in the thermal conductivity lies near the temperature at which the heat capacity exhibits a boson peak.  Is this a coincidence?

The answer to the first conundrum follows naturally from our result that there is a frequency regime of small and constant diffusivity.  As shown in Sec.~\ref{ac}, the plateau in the diffusivity above $\omega_{d}$ gives rise to a linear increase in thermal conductivity above $k_BT_{d} \sim \hbar \omega_{d}$.   Similarly, the vanishing of the density of states at $\omega_{max}$ leads to saturation of $\kappa(T)$ at high $k_B T> \hbar \omega_{max}$.  Thus, the transport crossover frequency sets the high-temperature limit of the plateau in $\kappa(T)$, while the maximum allowed frequency sets the temperature at which $\kappa(T)$ saturates to its final high-temperature value.
   
The answer to the second conundrum follows directly from the observation that $\omega_{d} \approx \omega^*$, since the high-temperature limit of the plateau in $\kappa(T)$ is $k_BT_{d} \sim \hbar \omega_{d}$ and the boson peak temperature is $k_B T_{BP} \sim \hbar \omega^*$.  Therefore, $T_{d} \approx T_{BP}$.

One might object that our results were obtained for a special system in which spheres interact via finite-ranged repulsions.   It has been argued theoretically \cite{matthieu2} and shown numerically \cite{Xu07} that packings of spheres interacting via Lennard-Jones interactions, which are attractive at long distances, behave much as compressed packings of repulsive spheres and that the boson peak frequency shifts upwards with increasing density in such systems.  Indeed, it was found that the $\omega^*$ is determined primarily by the repulsive interactions that come into play because the systems are held at high densities by the attractions.  Thus, even systems with attractions display a boson peak corresponding to the onset of anomalous modes, which should lead to a transport crossover as well.

It has also been shown that packings of ellipsoids \cite{ellipsoids} display boson peaks corresponding to the onset of anomalous modes that are similar in character to those for spheres.  For ellipsoids, the boson peak frequency is controlled by much the same physics as for spheres; it depends on the coordination number as in Eq. (\ref{eq:do3a}).

Another class of systems, network glasses, appears at first glance to be very different from our model repulsive sphere packings.  However, the unstressed models that were the main focus of this paper can be described as points (corresponding to the centers of spheres) connected by unstretched springs.  Such networks bear some resemblance to network glasses, which are held together by covalent attractions.  We include only central forces, but the counting of constraints has been shown to be key to network glasses with bond-bending forces as well~\cite{Phillips,Thorpe}.   The connection between harmonic spring networks and covalent network glasses has been discussed in more detail in Refs.~\cite{matthieu2} and ~\cite{matthieu3}.

In summary, our results suggest that two generic features of amorphous solids, the rise of the thermal conductivity with temperature above a plateau and the coincidence of the plateau temperature with the boson peak temperature, can be viewed as echoes of Point J.  This transition controls the onset frequency of anomalous modes with constant, minimal diffusivity in the manner of a critical point.
  
\begin{acknowledgments}
  We thank W. Ellenbroek, R. D. Kamien, T. C. Lubensky, Y. Shokef and T. A. Witten for
  helpful discussions.  This work was supported by DE-FG02-05ER46199 (AJL, NX
  and VV), DE-FG02-03ER46088 (SRN and NX) , NSF-DMR05-47230 (VV), and
  NSF-DMR-0213745 (SRN).

\end{acknowledgments}

\end{document}